\documentclass{article}
\usepackage{spconf,amsmath,graphicx}

\usepackage{tcolorbox}
\usepackage{times}
\usepackage{multirow}
\usepackage{amsfonts}
\usepackage{epsfig}
\usepackage{amsmath}
\usepackage{amssymb}
\usepackage[nolist]{acronym}
\usepackage[english]{babel}
\usepackage{cite}
\usepackage{color}
\usepackage{stfloats}
\usepackage{psfrag}
\usepackage{algorithm}
\usepackage{algorithmic}
\usepackage{subfigure}
\usepackage[algo2e]{algorithm2e} 
\usepackage{algorithmic}
\usepackage{hyperref}

\usepackage{adjustbox}
\usepackage{gensymb}
\usepackage{textcomp}

\newcommand{\ie}[0]{\textit{i.e.},~}

\newfloat{ilp}{ht}{aux}
\floatname{ilp}{Integer Linear Programming}

\title{Spherical clustering of users navigating 360$^\circ$ content}
 \name{Silvia Rossi$^{\star}$ \qquad Francesca De Simone$^{\dagger}$ \qquad Pascal Frossard$^{\ddagger}$ \qquad Laura Toni $^{\star}$ }
 	\address{$^{\star}$ Department of Electronic \& Electrical Engineering, UCL, London (UK)\\
 								$^{\dagger}$ DIS, Centrum Wiskunde \& Informatica, The Netherlands\\
 								$^{\ddagger}$ LTS4, \'Ecole Polytechnique F\'ed\'erale de Lausanne (EPFL), Switzerland\\
E-mails:					\fontsize{9}{9}\selectfont\ttfamily\upshape			  \{s.rossi, l.toni\}@ucl.ac.uk, 							F.De.Simone@cwi.nl, pascal.frossard@epfl.ch }
\thanks{This work has been supported by Royal Society under grant IES$\setminus$R1$\setminus$180128 and by Adobe under Academic Donation scheme.}

\begin{document}
	
	\maketitle
	\ninept   
	\begin{figure}[b]
		\begin{acronym}
			\acro{AR}{Augmented Reality}
			
			\acro{CDN}{Content Delivery Network}	
			\acro{CMP}{Cube Map Projection}
			\acro{CNN}{Convolutional Neural Network}
			\acro{CWI}{Centrum Wiskunde \& Informatica}
			
			\acro{DASH}{Dynamic Adaptive Streaming over HTTP}
			\acro{DOF}[3-DOF]{Degrees-Of-Freedoms}
			
			\acro{ERP}{Equirectangular Projection}
			
			\acro{FoA}{focus of attention}
			\acro{FoV}{field of view}
			\acro{FPR}{False Positive Rate}
			\acro{FSM}{fused saliency maps}
			
			\acro{GBVS}{Graph-Based Visual Saliency}
			
			\acro{HAS}{HTTP adaptive streaming}
			\acro{HQ}{High Quality}
			\acro{HMD}{head-mounted display}
			
			\acro{ILP}{integer linear programming}
			
			\acro{MCTS}{motion-constrained tile sets}
			\acro{MPEG}{Moving Picture Experts Group}
			\acro{MSE}{Mean Square Error}
			
			\acro{OMAF}{omnidirectional media application format}
			
			\acro{PSNR}{peak signal-to-noise ratio}
			
			\acro{QEC}{Quality Emphasis Center}
			\acro{QER}{Quality Emphasised Region}
			\acro{QoE}{Quality of Experience}	
			\acro{QoS}{quality-of-service } 
			
			\acro{ROC}{Receiver Operating Characteristic}
			\acro{RoI}{Region of Interest}
			
			\acro{SI}{Spatial Information}
			\acro{SRD}{Spatial Relationship Description}
			
			\acro{VoD}{Video on Demand}
			\acro{VR}{Virtual Reality}
			
			\acro{TCD}{Trinity College of Dublin}
			\acro{TI}{Temporal Information}
			\acro{TPR}{True Positive Rate}
			\acro{TSP}{Truncated Square Pyramid}
		\end{acronym}
		
				\parbox{\hsize}{\em
				Copyright 2019 IEEE. Published in the IEEE 2019 International Conference on Acoustics, Speech, and Signal Processing (ICASSP 2019), scheduled for 12-17 May, 2019, in Brighton, United Kingdom. Personal use of this material is permitted. However, permission to reprint/republish this material for advertising or promotional purposes or for creating new collective works for resale or redistribution to servers or lists, or to reuse any copyrighted component of this work in other works, must be obtained from the IEEE. Contact: Manager, Copyrights and Permissions / IEEE Service Center / 445 Hoes Lane / P.O. Box 1331 / Piscataway, NJ 08855-1331, USA. Telephone: + Intl. 908-562-3966.
		%
		}
	\end{figure}
	
	\begin{abstract}
In Virtual Reality (VR) applications,  understanding how users explore
the omnidirectional content is important  to optimize
content creation, to develop user-centric services,
or even to detect disorders in medical applications. Clustering users based on their common navigation patterns is a first direction to understand users behavior.  However, classical clustering techniques fail in identifying this common paths, since they are usually focused on minimizing a simple distance metric. In this paper, we argue that minimizing the distance metric does not necessarily guarantee to identify users that experience similar navigation path in the VR domain. Therefore, we propose a graph-based method to identify clusters of users who are attending the same portion of the spherical content over time. The proposed solution takes into account the spherical geometry of the content and aims at clustering users based on the actual overlap of displayed content among users. Our method is tested on real VR user navigation patterns. Results show  that  our solution leads to clusters in which at least 85\% of the content displayed by one user  is shared among the other users belonging to the same cluster.


	\end{abstract}

\keywords{Virtual Reality, 360$^\circ$ video, user behaviour analysis, data clustering}
		
	\section{Introduction }        
\ac{VR} systems are expected to become wide spread in a near future, with applications in a variety of fields, ranging from entertainment to e-healthcare. These systems involve omnidirectional (i.e., 360$^\circ$) videos, which are visual signals defined on a virtual sphere, depicting the 360$^\circ$ surrounding scene. 
The viewer, virtually positioned at the centre of the sphere, can navigate the scene with three \ac{DOF}, i.e., yaw, pitch and roll, by rotating his head and changing his viewing direction. This interactive navigation is typically enabled by a \ac{HMD}, which renders at each instant in time only the portion of the spherical content attended by the user, i.e., the \textit{viewport}. 

Understanding how users explore the VR content is important in order to optimize content creation~\cite{Serrano2017} and distribution~\cite{rossi2017navigation,corbillon2017viewport,petrangeli2017http,fan2017fixation,Yu2015}, develop user-centric services~\cite{Broeck2017,Sitzmann2018}, and even for medical applications that use VR to study psychiatric disorders~\cite{srivastava2014virtual}. 
In the last few years, many studies have appeared collecting and analysing the navigation patterns of users watching VR content~\cite{Yu2015,Upenik2017,hu2017head,wu2017,corbillon2017,lo2017360,fremerey2018avtrack360, Singla2017,Sitzmann2018}. 
Most studies build content-dependent \textit{saliency maps} as main outcome of their analysis, which compute the most probable region of the sphere attended by the viewers, based on their head or eye movements~\cite{Duchowski2002, Yu2015, david:2018:DHE:3204949.3208139, Upenik2017, Rai2017}. Some studies also provide additional quantitative analysis based on metrics, such as the average angular velocity, frequency of fixation, and mean exploration angles
~\cite{Sitzmann2018, corbillon2017}. 
Models to predict future saliency maps have also been proposed~\cite{Bogdanova2008,Bogdanova2010,xu2017modeling}.
Nevertheless, none of these studies performs clustering of the navigation patterns, i.e., none provides quantitative data indicating how many groups of 
users consistently share the same behaviour over time, by attending a significantly overlapping portion of the 360$^\circ$ content. This information can be useful in order to improve the accuracy and robustness of algorithms  predicting users navigation patterns. A proper clustering could also be useful   to refine user-centric distribution strategies, where for example different groups of users might be served with high quality content in the different portions of the sphere  that will be more likely attended by the viewers.

To the best of our knowledge,  studies identifying clusters for omnidirectional content delivery have appeared only recently \cite{xie2018cls,Gwendal2018}. User clustering is employed to identify the number of \acp{RoI} over time and to perform long-term prediction, associating to  each user   the future trajectory of the cluster that user belong to.
In \cite{xie2018cls}, the viewport center, i.e., the viewing direction of each user at each instant in time, is considered as a point on the equirectangular planar representation of the spherical content. These points on the plane are then clustered based on their Euclidean distance, which unfortunately ignores the actual spherical geometry of the navigation domain. Conversely, in~\cite{Gwendal2018} 
each user navigation pattern is modelled as independent trajectories in roll, pitch, and yaw angles, 
and spectral clustering is then applied. While it is efficient in discovering general trends of users' navigation, this clustering methodology might fail to identify clusters that are consistent in terms of actual overlap between viewports displayed by different users. It means that users in the same cluster do not necessarily consume the same portion of content.  At the same time, this consistency 
 needs to be guaranteed for clustering methods to be used for prediction purposes or for implementing accurate user-based delivery strategies. 

The goal of this paper is to propose a novel clustering strategy able to detect meaningful clusters in the spherical domain. 
We consider as meaningful cluster a set of users \emph{attending the same portion of spherical content} at a given time instant or over a series of frames.
This implies that the overlap between the viewports of all users in a cluster must be substantial.  
%
With this goal in mind, first   we define a metric to 
quantify the geometric overlap between two viewports on the sphere (Section II).
Then, we use this metric to build a graph whose nodes are the centers of the viewports associated to different users. Two nodes are connected only if the two corresponding viewports have a significant overlap (Section III). Finally, we propose a clustering method based on the Bron-Kerbosch (BK) algorithm \cite{bron1973algorithm} to identify clusters that are cliques, i.e.,  sub-graphs of inter-connected nodes (Section III).  
Results demonstrate the consistency of the proposed clustering method in identifying clusters where the overlap between the portions of the spherical surface corresponding to different viewports is higher than in state-of-the-art clustering (Section IV). 
 %
In summary, the main contribution of this paper is to propose a clustering algorithm that $i)$ considers the spherical geometry of the data, $ii)$ identifies clusters in which there is a consistent and significant 
geometric overlap between the portions of spherical surface corresponding to viewports attended by different users (by imposing that clusters are cliques), $iii)$ can be applied to a single frame or to a series of frames. 
This is a useful new tool to improve the accuracy of user's navigation prediction algorithms and user-dependent VR content delivery strategies, such as those proposed in~\cite{xie2018cls,Gwendal2018}.
	\section{Geodesic distance as proxy of viewport overlap } 
	\label{sec2}
Our goal is to identify clusters of users who are displaying the same portion of spherical content within a frame or over a series of consecutive frames. We derive a similarity metric that reliably quantifies how similar the portions attended by two users are. 
More specifically, each user attends a portion of the spherical surface. This is the projection on the spherical surface  of a plane tangent to the sphere (i.e., \textit{viewport}) in the point that identifies the user's viewing direction (\textit{center of the viewport})
\footnote{Without loss of generalization, we consider a scenario in which the viewports of all users have the same horizontal and vertical field of view.}.
The overlap between the viewports attended by two users at an instant in time is a clear indicator of how  similar users are with respect to their displayed  viewports.  For example,  an overlap equal to the area of the viewport corresponds to two users attending exactly the same portion of visual content. The geometric overlap could be analytically computed, knowing the rotation associated to each user head's position (i.e., roll, pitch, and yaw) and the horizontal and vertical fields of view that define the viewport. However, this is non trivial. 
Thus, we propose the simple and straightforward solution of using the \textit{geodesic distance} between two viewport centres  as a proxy for the viewport overlap. 

\begin{figure}[t]
		\centering
		\subfigure[$87\%$ overlap]{
			\includegraphics[width=0.15\textwidth]{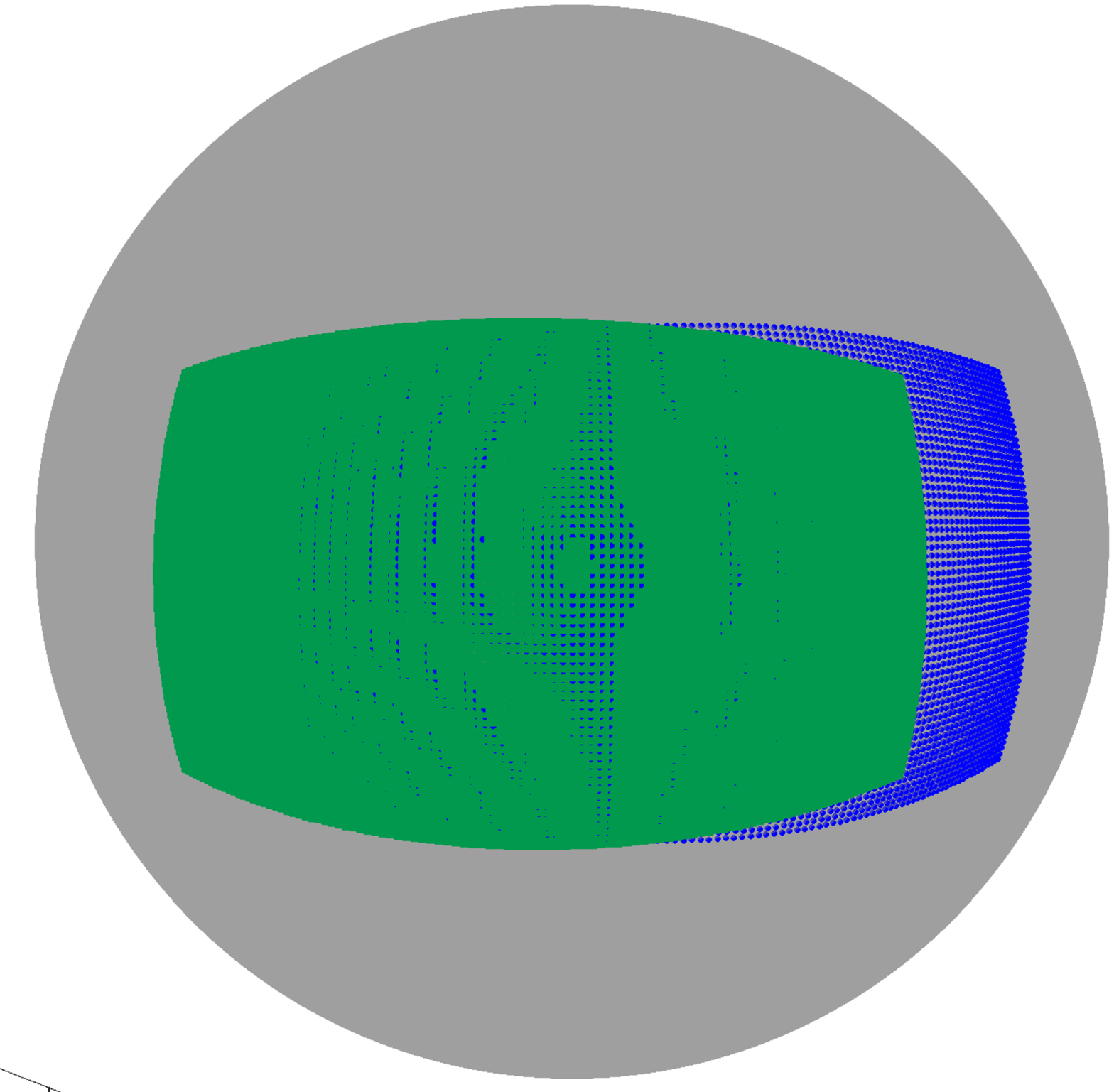}}
			\hfil		
		\subfigure[$58\%$ overlap ]{
			\includegraphics[width=0.15 \textwidth]{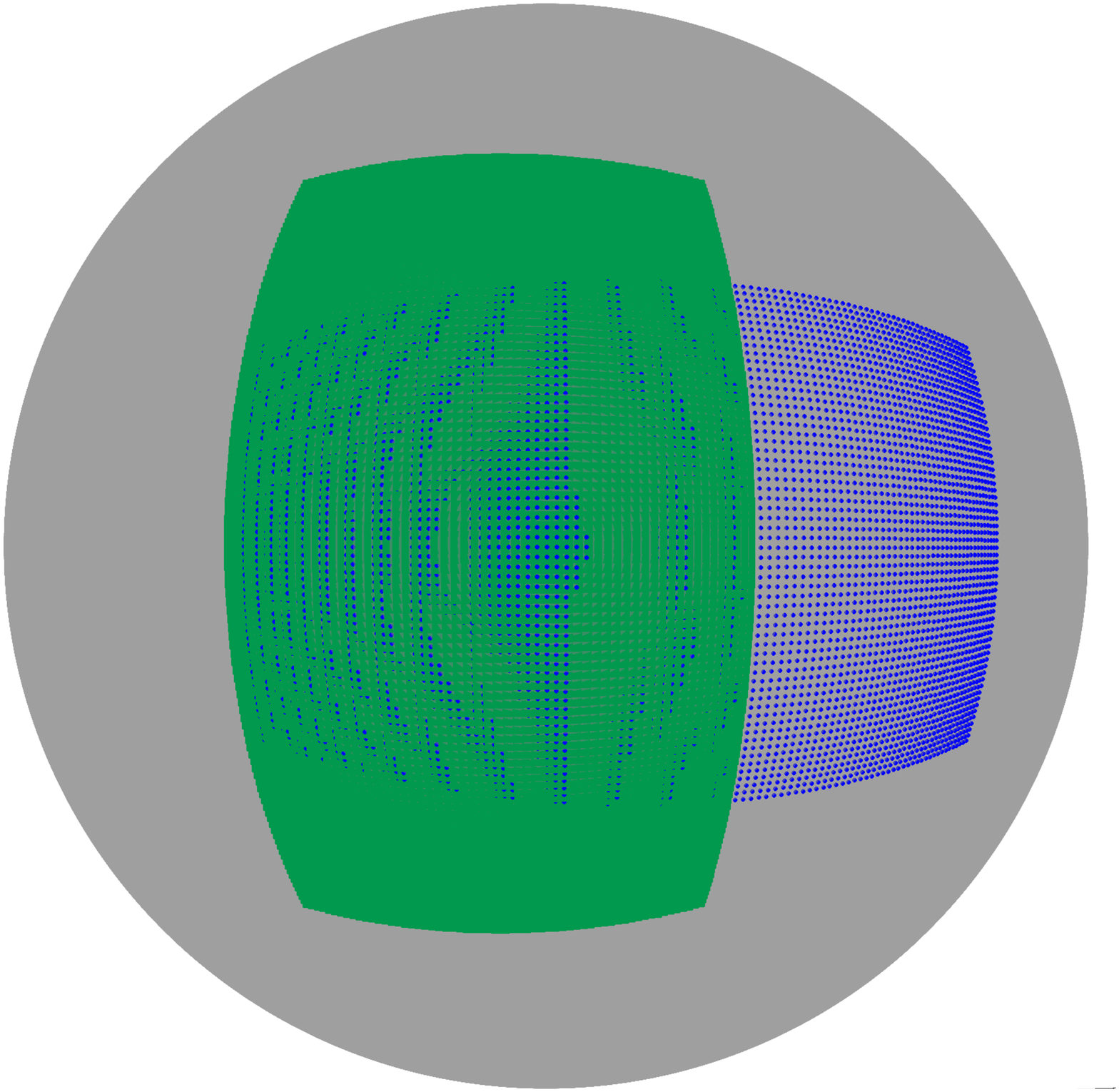}}		
		\caption{\label{fig_ch04:VPs_intersec} Viewports (in green and blue) with ${\pi}/{10}$ centre distance.  (a) viewports are aligned with an overlap of $87\%$, (b) one viewport is rotated by ${\pi}/{2}$ resulting an overlap of  $58\%$.}
	\end{figure}

\begin{figure} 
	\centering
		\includegraphics[width=0.40\textwidth]{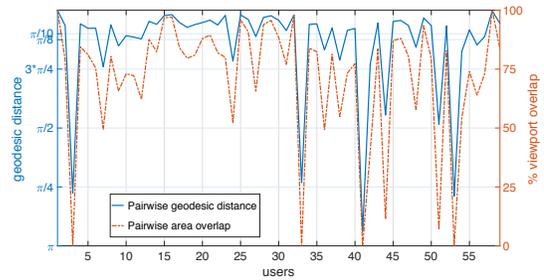}
	\caption{Comparison between pairwise geodesic distance and viewport overlap in one frame of video Rollercoaster from \cite{corbillon2017}. } 
	 \label{fig:ComparisonUser1}
\end{figure}

By  geodesic distance we denote  the length of the shortest arc connecting the viewport centers on the sphere.
Such distance is clearly an approximation of the actual area overlap: it does not account for the three degrees of freedom of the user's head rotation, which define the exact viewport. As a result, viewports whose centers have the same geodesic distance could correspond to a different viewport overlap (example in Figure~\ref{fig_ch04:VPs_intersec}).
Nevertheless, the smaller the distance between viewport centers, the smaller the approximation error with geodesic distance. As an example, Figure~\ref{fig:ComparisonUser1} shows the pairwise geodesic distance (in blue) and the pairwise area overlap (in red) between the viewport attended by one user and those of 58 other users, for a frame of a video sequence, extracted from the public dataset proposed in \cite{corbillon2017}. 
The correlation between the two metrics is evident: if the overlap is high, the geodesic distance between the two viewport centres is low. 
Particularly, a viewport area overlap larger than 75\% of the viewport area corresponds to a geodesic distance smaller than ${3\pi}/{4}$.  We are therefore interested in identifying a threshold value below which the geodesic distance is a robust proxy of the viewports overlap.  

\begin{figure}[t]
	\centering
		\includegraphics[width=0.40\textwidth]{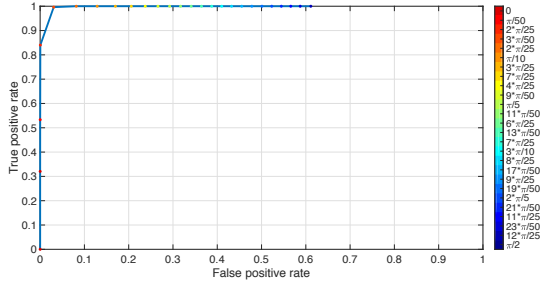} 
	\caption{\ac{ROC} curve to evaluate optimal $G_{th}$ considering all video in database \cite{corbillon2017} and $O_{th}=80\%$ \label{fig:ROC} .} 
\end{figure}

To empirically define this  threshold, we built the \ac{ROC}  curve as follows.   We assume that two users are attending   the same portion of   content if their viewports overlap by at least $O_{{th}}$ of the total viewport area. We then define a threshold value for the geodesic distance $G_{th}$ such that  users are neighbours if their geodesic distance is below threshold. Anytime users are neighbors but their overlap is less than $O_{{th}}$, we experience a false positive. Conversely, a true positive is experienced if users that are neighbors also experience an overlap equal or higher than $O_{{th}}$. 
Equipped with these definitions, we can compute the ROC  by considering all the videos and user' navigation patterns included in the dataset described in \cite{corbillon2017}.
Figure~\ref{fig:ROC} shows the curve obtained in our scenario with $O_{th}=80\%$.
On the x axis of the ROC curve there is the \ac{FPR} i.e., probability to have a wrong classification over the number of actual negative events. This rate should be as small as possible. On the contrary, the \ac{TPR} on the y axis represents the probability to correctly classify an event.
The best value of geodesic distance is ${\pi}/{10}$ since it corresponds to a \ac{TPR} value equal to 1, which in our application means a sure identification of viewports with an overlap of at least 80\% based on the geodesic distance between their centers. 
Therefore, in the following  we assume $G_{th} = {\pi}/{10}$ as a suitable threshold to robustly approximate the area overlap between two viewports by means of the geodesic distance between their centers.


\section{Clique-Based Clustering Algorithm}

We now describe the proposed clustering algorithm, aimed at identifying clusters of users having a common viewport overlap.
We model the evolution of users' viewports over a time-window $T$, i.e., a series of consecutive frames, as a set of graphs $\{ \mathcal{G}_t\}_{t=1}^T$. Each unweighted and undirected  graph $\mathcal{G}_t = \{\mathcal{V}, \mathcal{E}_t, \text{W}_t \}$ represents  the set of users\footnote{Without loss of generality, we assume that the set of users does not change over time. This covers also cases in which users' devices are not synchronized in the acquisition time, as users' positions are usually interpolated to create a synchronized dataset. } viewports at a particular instant $t$, where $\mathcal{V}$ and $\mathcal{E}_t$ denote the node and edge sets of $\mathcal{G}_t$. Each node in  $\mathcal{V}$ corresponds to a user interacting with the 360$^\circ$ content. Each edge in $\mathcal{E}_t$ connects neighbouring nodes, where two nodes are neighbours if the geodesic distance between the viewport centers associated to the users represented by the nodes is lower than $\mathcal{G}_t$, as defined in Section II. The binary matrix $\text{W}_t$ is the adjacency matrix of $\mathcal{G}_t$, with $w_t(i,j)=1$ if the geodesic distance between  the two viewport centres of users $i$ and $j$ at time $t$ is below a threshold. More formally:
\begin{equation}
	w_t(i,j) = 	 \begin{cases} 
	0,    & \mbox{if } g(i,j) \leq G_{th} \\ 
	1, & \text{otherwise} \\
	\end{cases} 
\end{equation}
where $g(i,j)$ is the geodesic distance between the viewport centres of 
users $i$ and $j$ and $G_{th}$ is thresholding value, discussed in Section II. Note that the clique-based clustering algorithm that we present in the following gets in input  binary adjacency matrices. Hence, $\text{W}_t $ is binary.

Looking at the graphs over time $\{ \mathcal{G}_t\}_{t=1}^T$, we are interested in clustering users based on their trajectories within a time window $T$.  
Similarly to other clusters of trajectories \cite{atev2010clustering}, we derive an affinity matrix $\text{A}$ that will be the input to our clustering algorithm, with
  \begin{equation}\label{eq:affinity}
	a(i,j) = \mathcal{I}_D	\left(\sum_{t=1}^T w_t(i,j) \right)
\end{equation}
where $\mathcal{I}_D(x)=1$ if $x\geq \tau$   and $0$ otherwise. 
This means that in the final graph two nodes, representing two users, are neighbours, i.e., connected by an edge, only if the corresponding viewports have a significant overlap in $D$ instants over $T$.
In the case of $\tau=T$, we obtain $a(i,j) =  \mathcal{I}_T \left(\sum_{t} w_t(i,j) \right) = \prod_{t} W_t$, and users' viewport centers need to be \emph{always} at a distance below threshold $\mathcal{G}_t$. This condition is however too constraining, therefore we introduce the threshold value $\tau$.

\begin{figure}[t]
		\centering
		\includegraphics[width=0.4\textwidth]{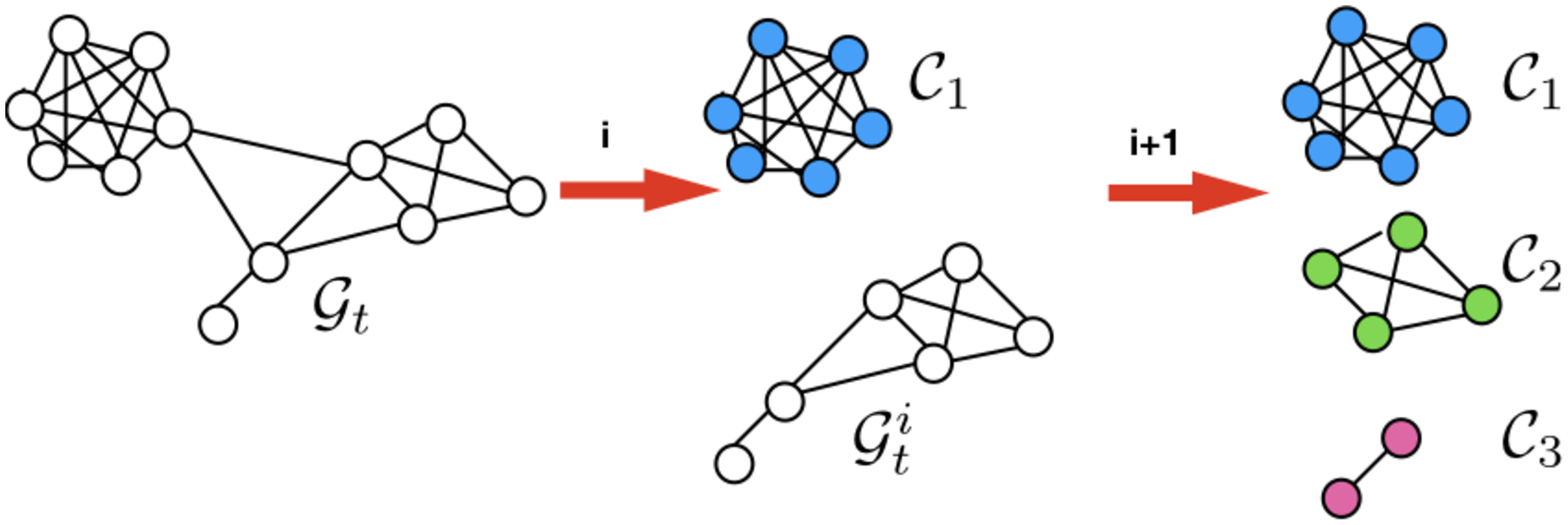} 
		\caption{\label{fig:CliqueClustering} Graphical example of the proposed clique clustering.}
	\end{figure}

The goal of our clustering algorithm is to identify groups of users that are consistently attending the same portion of the spherical surface. 
	To ensure that all users belonging to a cluster are attending the same area, they all need to be neighbors (i.e., $a(i,j)=1$ for all pairs of users $i$ and $j$ in the cluster). Therefore, we propose a \emph{clique-based clustering}. 
	In graph theory, a set of nodes all connected to each other is called a \emph{clique}.  
	A clique perfectly matches with the definition of cluster needed in our application, which identifies a set of users all having significant pairwise viewport overlaps, thus attending a common portion of video. 
	We consider the \emph{Bron-Kerbosch (BK) algorithm} \cite{bron1973algorithm} to find all \textit{maximal cliques} present in our graph (i.e., the most populated sub-graphs forming cliques). 
	However, maximal cliques identified by the BK algorithm can intersect, i.e., one user can belong to more than one clique. Conversely, we are interested in identifying disjoint sets\footnote{Clusters should be disjoint for most content-delivery applications.
	For example, if clusters are used for prediction, each user must belong only to one cluster.}.
	Hence, our clustering method consists of 
iterations of BK instances, as depicted in Figure~\ref{fig:CliqueClustering}. 
We initialize the clustering method by evaluating the affinity matrix from Eq.~\eqref{eq:affinity}. Then, we perform the following steps (Algorithm~\ref{Algorithm_CliqueCl}):
\begin{enumerate}
	\item Maximal cliques in the graph are detected by the BK algorithm. 
	\item  Among the resulting cliques, only the most populated one (i.e., the one with the highest cardinality) is kept as a cluster. 
	\item A new affinity matrix is built,  eliminating the entries corresponding to the elements of the cluster identified  in Step 2.
	\end{enumerate}
	These three steps are repeated until all nodes are assigned to clusters. It is worth mentioning that this iterative selection does not guarantee optimal clusters (i.e., clusters with maximal joint overlap among the viewports of users belonging to a cluster). However, $i)$ it imposes viewport overlap among users within a cluster, $ii)$ it identifies highly populated clusters, which can be translated in reliable trajectories/behaviours shared among users. 
	
	\begin{algorithm}[t]
	\begin{algorithmic} 
		\caption{Clique-Based  Clustering }\label{Algorithm_CliqueCl}
  		%
		\STATE \textbf{Input:} \hspace{0.4cm} $\{ \mathcal{G}_t\}_{t=1}^T$, D\\
		\STATE\textbf{Output: } \hspace{0.07cm}$K, \pmb{Q} = [Q_1,...,Q_K]$  \\
		\STATE\textbf{Init: } \hspace{0.07cm}$i=1$, $\text{A}^{(1)}=\mathcal{I}_D(\sum_tW_t),  \pmb{Q}=[\{\emptyset\}, \ldots, \{\emptyset\}]$ \\
	  		 
		\Repeat{ $A^{(i)}$ is not empty  }{
			\STATE  $\pmb{\mathcal{C}} = [\mathcal{C}_1,..., \mathcal{C}_{L}] \leftarrow   KB(\text{A}^{(i)})$ \\
 			\STATE   $l^{\star} =  \text{arg }\max_l |\mathcal{C}_l|$\\
 			\STATE  $ {Q}_i =   \mathcal{C}_{l^{\star}}$\\
			\STATE $\text{A}^{(i+1)} = \text{A}^{(i)}(\pmb{\mathcal{C}}  \setminus  \mathcal{C}_{l^{\star}})$
 			\STATE $i$ $\leftarrow$ $i+1$
		}
		\STATE $K=i-1$
		\end{algorithmic}
	 	\end{algorithm}

\section{Experimental results}
The proposed clustering algorithm is compared to state-of-the-art solutions, namely the \emph{Louvain method}  \cite{blondel2008fast}, the  \emph{K-means clustering} \cite{hartigan1979algorithm} and the clustering of \ac{VR} trajectories proposed in \cite{Gwendal2018} (labelled ``SC").  
We use the geodesic distance between viewport centers as distance metric in all algorithms.
Moreover, in the  \emph{K-means clustering}, the number of clusters $K$ is imposed as the value achieved by the Louvain method (labelled ``K-means 1"), as well as the $K$ value obtained from our proposed clustering  (labelled ``K-means 2"). The proposed implementations have been made publicly available~\footnote{{https://github.com/LASP-UCL/spherical-clustering-in-VR-content}.}.
 We test these algorithms on two 1-minute long video sequences (Rollercoaster and Timelapse), which have been watched  by $59$ users whose navigation paths are publicly available~\cite{corbillon2017}. Rollercoaster has one main \ac{RoI} (i.e., the rail) while in Timelapse, there are many fast moving objects (e.g., buildings, people) along the equator line. 

{\textbf{Frame-based Clustering.}}
First, we consider frame-based clustering, in which users are identified by their viewport centers at one given frame. Table \ref{table} reports results in terms of number of clusters ($K$), mean viewport overlap computed within each cluster composed by at least three users, and viewport overlap within the most populated cluster, that we refer to as the main cluster. The viewport overlap within a cluster is the joint overlap across all users' viewports in the cluster. The mean overlap is computed by averaging the viewport overlap of all clusters with at least three users identified at a given frame. 
In Table \ref{table}, we also provide 
the percentage of users covered by clusters. The  proposed algorithm  always ensures the highest viewport overlap (on average always over 50\%) with respect to the other methods. This is due to the implicit constraint that is imposed by the clique-based detection of the clusters. This constraint leads to the identification of clusters that are populated and yet meaningful (i.e., with large viewport overlap among users). For example, in Rollercoaster at frame $40s$, our algorithm identifies a main cluster grouping $35\%$ of the population with a viewport overlap of $58.33\%$. This is much higher than the overlap of $24.20\%$ ($0\%$) in the main cluster identified by the Louvain (K-means) method. 
Beyond the accuracy, another important parameter is the percentage of 
the population that is covered by clusters with a significant number of users. These clusters are the most useful ones to allow predictions.
For instance in Timelapse at frame 50$s$, our method identifies a large number of clusters (29), which also includes single users clusters. 
Nevertheless, half of the population (51.70\%) belongs to clusters with more than $3$ users with high value of joint overlap (71.40\%).
\begin{table*}[]
	\centering
	\resizebox{1\textwidth}{!}{%
		\begin{tabular}{cc||c|c|c|c||c|c|c|c|}
			\cline{3-10}
			&                                  & \multicolumn{4}{c|}{\textbf{ROLLERCOASTER}}                                                                                                                                                                                                                                                                                                          & \multicolumn{4}{c|}{\textbf{TIMELAPSE}}                                                                                                                                                                                                                                                                                                            \\ \cline{3-10} 
			&                                  & \textit{Louvain method}                                                                                & \textit{Clique Clustering }                                                                         & \textit{K-Means 1   }                                                                     & \textit{K-Means 2 }                                                                   & Louvain                                                                              & \textit{Clique Clustering }                                                                        & \textit{K-Means 1}                                                                      & \textit{K-Means 2}                                                                       \\ \hline
			\multicolumn{1}{|c|}{\multirow{4}{*}{\begin{tabular}[c]{@{}c@{}} {\rotatebox[origin=c]{90}{\small \hspace{0.18cm} Fr. \texttildelow30s}}\end{tabular}}}  & K                                & 10                                                                                     & 15                                                                                  & 10                                                                                  & 15                                                                              & 13                                                                                   & 24                                                                                 & 13                                                                                & 24                                                                                 \\ \cline{2-10} 
			\multicolumn{1}{|c|}{}                                                                               & Mean Overlap Cl.(\% user \textgreater 3)                   & 38.90 \%    (84.75 \%)                                                                           & 62.50 \%           (76.30 \%)                                                                 & 53.95 \%      (93.20 \%)                                                                      & 48.10 \%      (94.90 \%)                                                                  & 46 \%       (89.70\%)                                                                         & 72.35\%      (56.90\%)                                                                      & 45.90\%   (96.50 \%)                                                                         & 51.50\%          (50\%)                                                                   \\ \cline{2-10} 
			\multicolumn{1}{|c|}{}                                                                               & 
			Main cl. overlap (\% users)
			& \begin{tabular}[c]{@{}c@{}}
				 26.70\% (44.10\%)\end{tabular}    & \begin{tabular}[c]{@{}c@{}}
				58.60\% (30.50\%)\end{tabular}   & \begin{tabular}[c]{@{}c@{}}
				48.30\% (19\% )\end{tabular}      & \begin{tabular}[c]{@{}c@{}}
				0\% (20.70\% )\end{tabular} & \begin{tabular}[c]{@{}c@{}}
				32.90\% (20.70\%) \end{tabular}   & \begin{tabular}[c]{@{}c@{}}
				69\% (12.10\% )\end{tabular}      & \begin{tabular}[c]{@{}c@{}}
				15\% (19\% )\end{tabular}         & \begin{tabular}[c]{@{}c@{}}
				23.50\% (13.80\%) \end{tabular} \\ \hline 
			\multicolumn{1}{|c|}{\multirow{4}{*}{\begin{tabular}[c]{@{}c@{}}{\rotatebox[origin=c]{90} {{\hspace{0.3cm}Fr. \texttildelow40s}}}\end{tabular}}} & K                                & 8                                                                                      & 15                                                                                  & 8                                                                                   & 15                                                                              & 18                                                                                   & 27                                                                                 & 18                                                                                & 27                                                                                 \\ \cline{2-10} 
			\multicolumn{1}{|c|}{}                                                                               & Mean Overlap Cl.(\% users \textgreater 3)            & 35.60\%  (89.83\%)                                                                                & 65.75\%   (76.30\%)                                                                           & 
			44.38\%    (100\%)                                                                          & 
			47.65\%  (84.75\%)                                                                              & 
			47.65 \%  (75.90\%)                                                                           & 
			72.95 \%  (77.60\%)                                                                         & 60.27\%  (96.55\%)                                                                          & 65.90\%   (84.50\%)                                                                           \\ \cline{2-10} 
			\multicolumn{1}{|c|}{}& Main cl. overlap (\% users)                & 
			\begin{tabular}[c]{@{}c@{}}
				24.20\% (45.80\%)\end{tabular}    & \begin{tabular}[c]{@{}c@{}}
				58.33\% (35.60\%)\end{tabular}   & \begin{tabular}[c]{@{}c@{}}
				0\% (30.50\%)\end{tabular}     & \begin{tabular}[c]{@{}c@{}}
				0\% (15.25\%)\end{tabular}  & \begin{tabular}[c]{@{}c@{}}
				51.80\% (20.70\%)\end{tabular} & \begin{tabular}[c]{@{}c@{}}
				63.70\% (17.24\%)\end{tabular}  & \begin{tabular}[c]{@{}c@{}}
				47.50\% (20.70\%)\end{tabular} & \begin{tabular}[c]{@{}c@{}}
				33.60\% (8.60\%)\end{tabular}  \\ \hline
			\multicolumn{1}{|c|}{\multirow{4}{*}{\begin{tabular}[c]{@{}c@{}}{\rotatebox[origin=c]{90} {\hspace{0.3cm}Fr. \texttildelow50s}}\end{tabular}}} & K                                & 8                                                                                      & 12                                                                                  & 8                                                                                   & 12                                                                              & 18                                                                                   & 29                                                                                 & 18                                                                                & 29                                                                                 \\ \cline{2-10} 
			\multicolumn{1}{|c|}{}                                                                               & Mean Overlap Cl.(\% users \textgreater 3)                    & 
			48.20\%  (89.80\%)                                                                              &
			 65.70\%   (86.45\%)                                                                           & 
			 43.50\%  (98.30\%)                                                                            &
			 55.30\%   (96.60\%)                                                                        & 
			 49.12 \%    (77.60\%)                                                                          & 71.40\%      (51.70\%)                                                                      & 48.36 \%    (87.90\%)                                                                      & 55.90 \%     (55.17\%)                                                                        \\ \cline{2-10} 
			\multicolumn{1}{|c|}{}  &Main cl. overlap (\% users) & 
			\begin{tabular}[c]{@{}c@{}}
				46.40\%(30.50 \%)\end{tabular} & \begin{tabular}[c]{@{}c@{}}
				59.90\% (57.70\%)\end{tabular}   & \begin{tabular}[c]{@{}c@{}}
				0\% (22.40\%) \end{tabular}       & \begin{tabular}[c]{@{}c@{}}
				0\% (15.25\%)\end{tabular}   & \begin{tabular}[c]{@{}c@{}}
				30.60 (22.40\%)\%\end{tabular}   & \begin{tabular}[c]{@{}c@{}}
				70.80\% (25.90\%)\end{tabular} & \begin{tabular}[c]{@{}c@{}}
				37\% (24.15\%)\end{tabular}  & \begin{tabular}[c]{@{}c@{}}
				62.71\% (17.24\%)\end{tabular}   \\ \hline
		\end{tabular}%
	}
	\caption{Clustering analysis of users in three selected frames from Rollercoaster (first half) and Timelapse (second half). In brackets, the percentage of covered population. }
	\label{table}
\end{table*}

\begin{figure}[t]
	\centering
	\subfigure[Rollercoaster video - T = 3 s. ]{
		\includegraphics[width=0.38\textwidth]{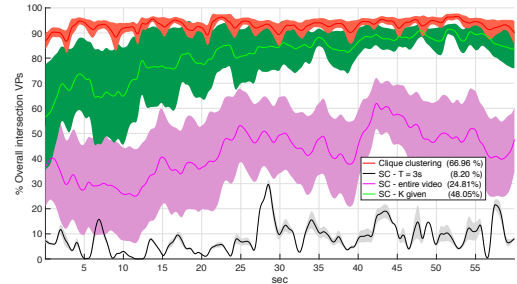}}
	\subfigure[Timelapse video - T = 3 s.]{
		\includegraphics[width=0.38\textwidth]{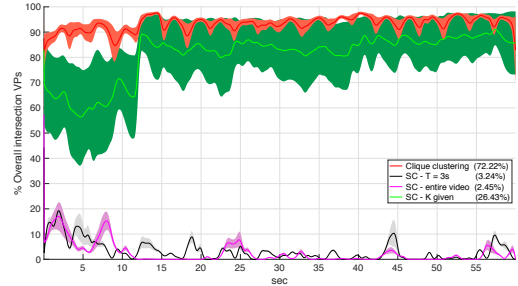}}
	\caption{\label{fig:AllVideo_OvInter_Var} Mean and variance of the joint overlap across clusters over time. In the legend, the mean value of joint viewport overlap of clusters with more than three users performed across the entire video.}
\end{figure}

 {\textbf{Trajectory-based clustering.} } Second, we test the proposed algorithm over a time-window with $T = 3s$ and $\tau = 1.8s $. 
 In this case we compare the proposed solution with algorithm SC~\cite{Gwendal2018}. 
 The algorithm SC is applied to trajectories spanning the entire video, as in~\cite{Gwendal2018}, as well as consecutive time windows of 3$s$. We also consider the case of SC in which the number of clusters $K$ is not evaluated from their affinity matrix
but it is imposed as the $K$ obtained from our solution. We label this clustering (``SC - K given"). Figure~\ref{fig:AllVideo_OvInter_Var} shows results in terms of overlap 
 among viewports clustered together in both Rollercoster (a) and Timelapse (b).
In more details, all users are clustered over consecutive time-windows of $T$ seconds each. 
 Then, for each frame the viewport overlap among all users within one cluster is evaluated and averaged across clusters. The mean overlap (solid line) and the variance (shaded area) is finally depicted in the figure.  Moreover, the mean value of joint overlap in clusters with more than three users across the entire video is showed in the legend. Our solution outperforms SC in terms of mean overlap but also in terms of variance. The latter shows the stability of our clustering method ensuring for each cluster a consistent overlap over time. Finally, the performance gain is significant also in terms of overlap in the most populated clusters (value provided in the legend). 

	\section{Conclusions}
In this paper,  we proposed a novel graph-based clustering strategy able to detect meaningful clusters, \ie group of users consuming the same portion of  a virtual reality   spherical content. First, we derived a geodesic distance threshold value to reflect the similarity among users and then we built a clique-based clustering based on this metric. 
%
Results  carried out on real VR  user navigation patterns show that the proposed method identifies clusters  with higher joint overlap than other state-of-the-art clustering methods. 
 Future works will focus on the application of our method in the framework of adaptive streaming of VR videos and for the prediction of user navigation patterns.
		\bibliographystyle{IEEEbib}
	\bibliography{reference}

	
\end{document}